\documentclass[aps,pra,twocolumn,english,preprint,10pt,nofootinbib]{revtex4}
\usepackage[T1]{fontenc}
\usepackage[latin9]{inputenc}
\usepackage{babel}
\usepackage{nccmath}

\usepackage{graphicx,amsmath,amssymb,color}
\usepackage[normalem]{ulem}
\usepackage{amsfonts}
\usepackage[toc,page]{appendix}
\usepackage{hyperref}
\usepackage{latexsym}
\usepackage{amsfonts}
\usepackage{algpseudocode}
\usepackage{amsthm}
\usepackage{mathrsfs}
\usepackage{natbib}
\usepackage{color,verbatim}
\usepackage{psfrag}
\usepackage[svgnames]{xcolor}

\preprint{IPPP/20/22}

\begin{document}

\title{Observing the fate of the false vacuum with a quantum laboratory}
\author{Steven Abel}
\email{s.a.abel@durham.ac.uk}
\address{Institute for Particle Physics Phenomenology, Department of Physics, Durham University, Durham DH1 3LE, U.K.}

\author{Michael Spannowsky}
\email{michael.spannowsky@durham.ac.uk}
\address{Institute for Particle Physics Phenomenology, Department of Physics, Durham University, Durham DH1 3LE, U.K.}

\begin{abstract}
 We design and implement a quantum laboratory to experimentally observe and study dynamical processes of quantum field theories. Our approach encodes the field theory as an Ising model, which is then solved by a quantum annealer. As a proof-of-concept, we encode a scalar field theory and measure the probability for it to tunnel from the false to the true vacuum for various tunnelling times, vacuum displacements and potential profiles. The results are in accord with those predicted theoretically,  showing that a quantum annealer is a genuine quantum system that can be used as a quantum laboratory. This is the first time it has been possible to experimentally measure instanton processes in a freely chosen quantum field theory. This novel and flexible method to study the dynamics of quantum systems can be applied to any field theory of interest. Experimental measurements of the dynamical behaviour of field theories are independent of theoretical calculations and can be used to infer their properties without being limited by the availability of suitable perturbative or nonperturbative computational methods. In the near future, measurements in such a quantum laboratory could therefore be used to improve theoretical and computational methods conceptually and may enable the measurement and detailed study of previously unobserved quantum phenomena.
 \end{abstract}

\maketitle

\section{Introduction}

Quantum field theories are the theoretical framework underlying the most fundamental description of nature.  Yet, studying the dynamics of those special classes of quantum field theories that occur in nature requires the use of either multi-body quantum systems, e.g. condensed matter systems, or the design of highly sophisticated high-energy experiments that probe the properties of quantum field theories when they manifest themselves as particles. So far no quantum lab has been devised that provides an experimental framework to study the quantum effects and dynamics of arbitrary field theories, that is theories in which the quantum numbers and interactions of quantum fields can be adjusted at will. We show that a quantum annealer acting on a generalised Ising model\footnote{D-Wave Systems \cite{LantingAQC2017} provides currently access to a quantum annealer with 2048 qubits and a more connected 5000 qubit machine in the future.} is exactly that -- a quantum lab for arbitrary field theories. Consequently, in the near future many theoretical calculations for quantum field theories could be replaced by quantum experiments, thereby overcoming computational or theoretical limitations, e.g. perturbative or non-perturbative computational methods, or high computational demands, e.g. in lattice calculations.

In this paper we utilise the method introduced in Ref.\cite{Abel:2020ebj} for encoding a general field theory on a quantum annealer, and show that it allows one to implement and observe truly quantum dynamical processes. We will focus on recreating and measuring the phenomenon of tunnelling in scalar field theories which is clear evidence for a quantum rather than a classical process. For a field theory that has $d=0$ or $d=1$ spacetime dimensions  this is equivalent to the measurement of the time-dependent quantum mechanical wave function as it attempts to reach the ground state of the system. From a more general field theory perspective, it means that we are able to measure nonperturbative decay processes that are described by instantons  \cite{Coleman:1977py, Callan:1977pt, 0521318270, Affleck:1980ac, Linde:1981zj, Garriga:1994ut, Calzetta:2001pp}.

We stress that our method is very flexible and can be used to probe all kinds of non-perturbative  processes. For example if one were to consider $d=4$ spacetime dimensions  it would be possible to observe instanton processes in conventional relativistic quantum field theories. In Yang-Mills Theories \cite{Belavin:1975fg} such objects are of profound importance, because they are the  primary explicit example of genuinely non-perturbative gauge field configurations, leading to a wealth of geometrical, topological and quantum effects with a fundamental impact on quantum dynamics. Instanton effects are important in the electroweak sector of the Standard Model (SM) and also in QCD, as well as in a broad variety of theoretical constructions ranging from Supersymmetric models, GUT theories, extra dimensions, to string theory and D-branes \cite{Vainshtein:1981wh, Affleck:1983rr, Witten:1995gx, Dorey:1999pd, Maldacena:2001xj, Dorey:2002ik}. 
While all these incarnations of instantons have been predicted, and there is little doubt about their existence in the Standard Model and their profound role in shaping the history of the early Universe, none of the relevant processes have been observed experimentally \cite{Ringwald:1994kr,Brooijmans:2016lfv, Ellis:2016dgb,Ringwald:2018gpv, Khoze:2019jta}. There are by contrast several specific condensed matter systems where 
such effects have been observed\footnote{In non-relativistic field theories describing spin systems of chiral magnets with a Dzyaloshinskii-Moriya intereaction \cite{Dzyaloshinsky, Moriya:1960zz} magnetic skyrmions and domain walls have been observed \cite{bogdanov,rossler}. And instantons can appear in such systems as composite solitons \cite{Hongo:2019nfr}. Further, the measurement of instantons in special condensed matter systems, see e.g. \cite{chudnovsky, stamp, Zhang}, have been reported.}. However, all these cases were constrained to the particular field theory in question. The aim of this work is to instead provide a 
framework for studying nonperturbative effects in {\it any} field theory of interest. 

This in principle allows one to check the calculation of nonperturbative phenomena by studying them experimentally. It may even be possible to observe new phenomena that have not yet been anticipated. For this study we will of course be limited by the hardware that is available to us, so the discussion is necessarily restricted to the simpler field theories that can exhibit instanton-like behaviour, namely the aforementioned $d=1$ scalar field theory. Nevertheless, within this theory we will be able to set-up a potential that we then manipulate by hand so that it develops a non-trivial vacuum structure that induces tunnelling. We believe this is the first time that it has been possible to implement instanton processes in a freely chosen quantum field theory and observe such phenomena experimentally. 

\section{Set-up for false vacuum decay}

It will be convenient for several practical reasons to set-up a physical system on the 
annealer that recreates quantum decay in a potential of the form 
\begin{equation}
U(\phi)=    \frac{3}{4}\tanh^{2}\phi-{k(t)}\,\text{sech}^{2}\left(c(\phi-v)\right)  ,\label{eq:potential}
\end{equation}
where $c,v$ are constants while $k$ is time-dependent, and
$\phi(t)$ is the field. Note that $\phi$ is the dimensionless object 
that we will define on the annealer. When required we will convert it into a 
dimensionful field 
$\eta$ by defining 
\begin{equation}
\phi \,=\, \eta /\eta_0\, ,
\end{equation} 
where $\eta_0$ is a constant. 
 In the $d=1$ field theory there are of course
no space dimensions, and at leading order it is isomorphic to quantum
mechanics (with $\phi$ playing the role of $x$). However the $d=1$
field theory formalism allows for particle creation and is the starting
point for generalisation to higher dimensions, as discussed in the introduction. 

\begin{figure}
\centering{}\includegraphics[scale=0.46]{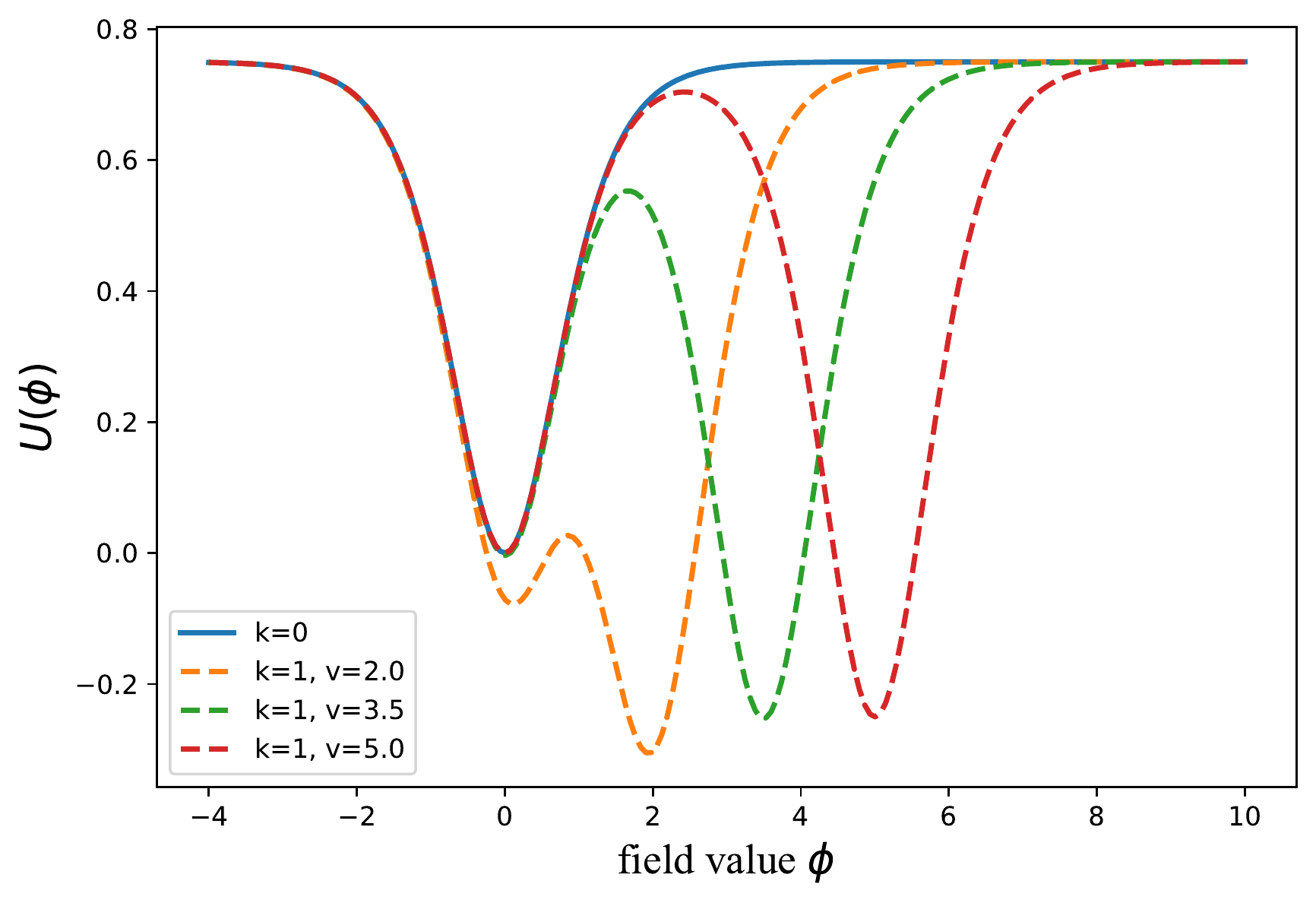}\caption{The double-P\"oschl-Teller potential well for different $k$ and $v$. The system is initialised
around $\phi=0$ and allowed to decay to the true minimum at $\phi\approx v$. \label{fig:PT-potential}}
\end{figure}

The first term in $U$ provides a potential-well around $\phi=0$ which in principle allows
the system to begin as a bound-state there. As mentioned this is one
of the benefits of annealers over discrete gate systems: in order
first to reach a ground state, a system has to dissipate. The $k$-term will then be turned on
adiabatically during the anneal in order to allow tunnelling into
the global minimum that forms at $\phi=v$. For this study we shall
mostly take $c=1$, so that the potential during the tunnelling period
will consist of equally sized potential wells. The potential is plotted in Fig.\ref{fig:PT-potential}
for $k=1$ and various values of separation parameter $v$. 

This function has several nice properties
for our purposes. One is that each individual well has the P\"oschl-Teller
$-\text{sech}^{2}\phi$ form, which can be solved. Moreover the potentials
around each minimum decay exponentially. This makes it possible to
``turn on'' the global true minimum by adjusting $k$ without significantly altering
the profile of the potential around the false minimum (unlike the
more commonly considered case of quartic potentials). Other useful
features of this choice will be discussed below when they become relevant.

We will begin the system with $k=0$, such that it falls into a P\"oschl-Teller ground state. Assuming that 
the completion of the potential into a $d=1$ field theory ultimately corresponds to the Schr\"odinger equation, 
the ground state (and its excited friends) in such a potential can be determined
using factorisation and ladder-operator methods (see for example \cite{Cevik:2016mnr,brown2018}).
In a theory where 
\begin{equation}
\label{eq:eigenvalues}
\frac{2m\eta_0^2}{\hbar^{2}} U  = \lambda(\lambda+1)\,\text{tanh}^2\phi ,
\end{equation}
the bound states are given by Legendre polynomials of the form $P_\lambda^\mu(\tanh\,\phi)$, and the ground state,
$P_\lambda^\lambda(\tanh\,\phi)$, is given by 
\begin{equation}
\psi_{0}(\phi)={\cal N}_0 \, \text{sech}^{\lambda}\phi ~,\label{eq:groundstate}
\end{equation}
where the normalisation constant is 
$${\cal N}^2_0 = \pi^{-\frac{1}{2}} {\Gamma(\lambda+1/2)}/{\Gamma (\lambda) }\, .
$$This state, which is our idealised starting state, has energy 
\begin{equation}
E_{0}=\frac{\hbar^{2}\lambda}{2m\eta_0^2}.
\end{equation}

We will not know {\it a priori} 
the value of  $$\gamma\stackrel{\rm def}{=} \hbar^2/2m\eta_0^2$$ in the effective  field theory induced on the annealer, and estimating it will essentially constitute our calibration. 
In order to do this we could for example multiply $U$ by a constant, $\alpha$ say, and by trial-and-error find a value for $\alpha$ that yielded a ground state wave function of the form $\psi_0 = \text{sech}(\phi) /\sqrt{\pi }$ corresponding to $\lambda=1/2$. According to \eqref{eq:eigenvalues} that value of $\alpha$ would be equal to  $\gamma$. However this is demanding to do (in terms of annealer time), and it is not always obvious which is the value of $\lambda$. We will instead determine an estimate for $\gamma$ in the effective  field theory by studying the ground state of the simple-harmonic-oscillator (SHO) potential, and fitting the wave-function to the ground state. Either way it is unavoidable that one must also determine  $\gamma$ as an empirical parameter.

Let us now consider the  tunnelling into the global minimum once
$k$ is turned on. 
The expected decay rate  can be computed using instanton methods. In
$d=1$ dimensional field theory this means writing the path
integral for the non-relativistic propagation of the physical field $\eta=\eta_0 \phi$ as a worldline
integral:
\begin{equation}
\langle\eta_{i}|\eta_{f}\rangle=\int_{\eta(0)=\eta_{i}}^{\eta(T)=\eta_{f}}\mathcal{D}\eta\,e^{-i\hbar^{-1}\int_{0}^{T}dt\left(\frac{1}{2}m\dot{\eta}^{2}-(U-E_0)\right)},
\end{equation}
where the path is between points $\eta_{i}$ inside and $\eta_{f}$
outside the barrier and $T$ is the time. As usual the integral is
dominated by the stationary phase contribution, but in order to evaluate
it efficiently we deform $t$ in the complex $t$ plane by making
a Wick rotation $t\rightarrow-it$ and use the Euclidean steepest-descent
contour instead: 
\begin{equation}
\langle\eta_{i}|\eta_{f}\rangle_{E}=\int_{\eta(0)=\eta_{i}}^{\eta(T)=\eta_{f}}\mathcal{D}\eta\,e^{-\hbar^{-1}\int dt\left(\frac{m\dot{\eta}^{2}}{2}+U-E_0\right)}.
\end{equation}
This describes the propagator from $\eta_{i}$ to the endpoint, but
we are most interested in the exponentially decaying part. The steepest
descent contour that determines it corresponds to the classical solution
of the Euclidean equation of motion $\eta_{cl}$ with endpoints at
$\eta_{+},\eta_{e}$, where $\eta_{e}$ is the escape point, namely
the point where $U=E_0$, with the quantum fluctuations providing pre-factors.
That is 
\begin{equation}
\delta S_{E}=0\implies m\ddot{\eta}=U_{\eta},
\end{equation}
which gives the usual classical solution 
\begin{equation}
\dot{\eta}_{cl}=\pm\sqrt{2(U-E_0)/m},
\end{equation}
corresponding to energy conservation for a ball rolling in the inverted
potential between turning points at $\eta_{+}$ and $\eta_{e}$. Substituting then gives 
the classical action
\begin{equation}
S_{E,cl}
\,\,=\,\,\int_{\eta_{+}}^{\eta_{e}}d\eta \sqrt{2m(U-E_0)}\,,
\end{equation}
and letting $\eta=\eta_{cl}+\delta\eta$ yields a quantum prefactor;
\begin{align}
\langle\eta_{i}|\eta_{f}\rangle_{E} & =\int\mathcal{D}\delta\eta\,e^{-\hbar^{-1}\int dt\left(\frac{m(\dot{\eta}_{cl}+\delta\dot{\eta})^{2}}{2}+U(\eta_{cl}+\delta\eta)-E_0\right)},\nonumber \\
 & =Ae^{-\hbar^{-1}S_{E,cl}},
\end{align}
with the decay rate $\Gamma=|\langle\eta_{i}|\eta_{f}\rangle_{E}|^{2}$
becoming 
\begin{equation}
\Gamma\approx e^{-2\hbar^{-1}S_{E,cl}}~.
\end{equation}
In principle these solutions should then be matched on to oscillating
solutions at the turning points, but these oscillating parts do not change
the decay rate. Thus regardless of the time $T$, the exponential
decay in the amplitude between points either side of the barrier will
be dominated by this saddle point approximation, as one would expect.
As mentioned the $d=1$ field theory is isomorphic to the Schr\"odinger
equation (SE) at leading order and indeed the same result can be obtained
using the WKB method. However the $d=1$ system actually includes
all the paraphernalia of field theory, including loop corrections,
particle pair production and so forth. In principle then it presents
a useful laboratory for testing both perturbative and nonperturbative
aspects of quantum field theory, and future generalisation of our
discussion to higher dimensions could be performed very straighforwardly
by including discretised space derivative terms. Only the limited
dimensions and connectivity of the annealer prevent us doing this. 

\begin{figure}
\centering{}\includegraphics[scale=0.5]{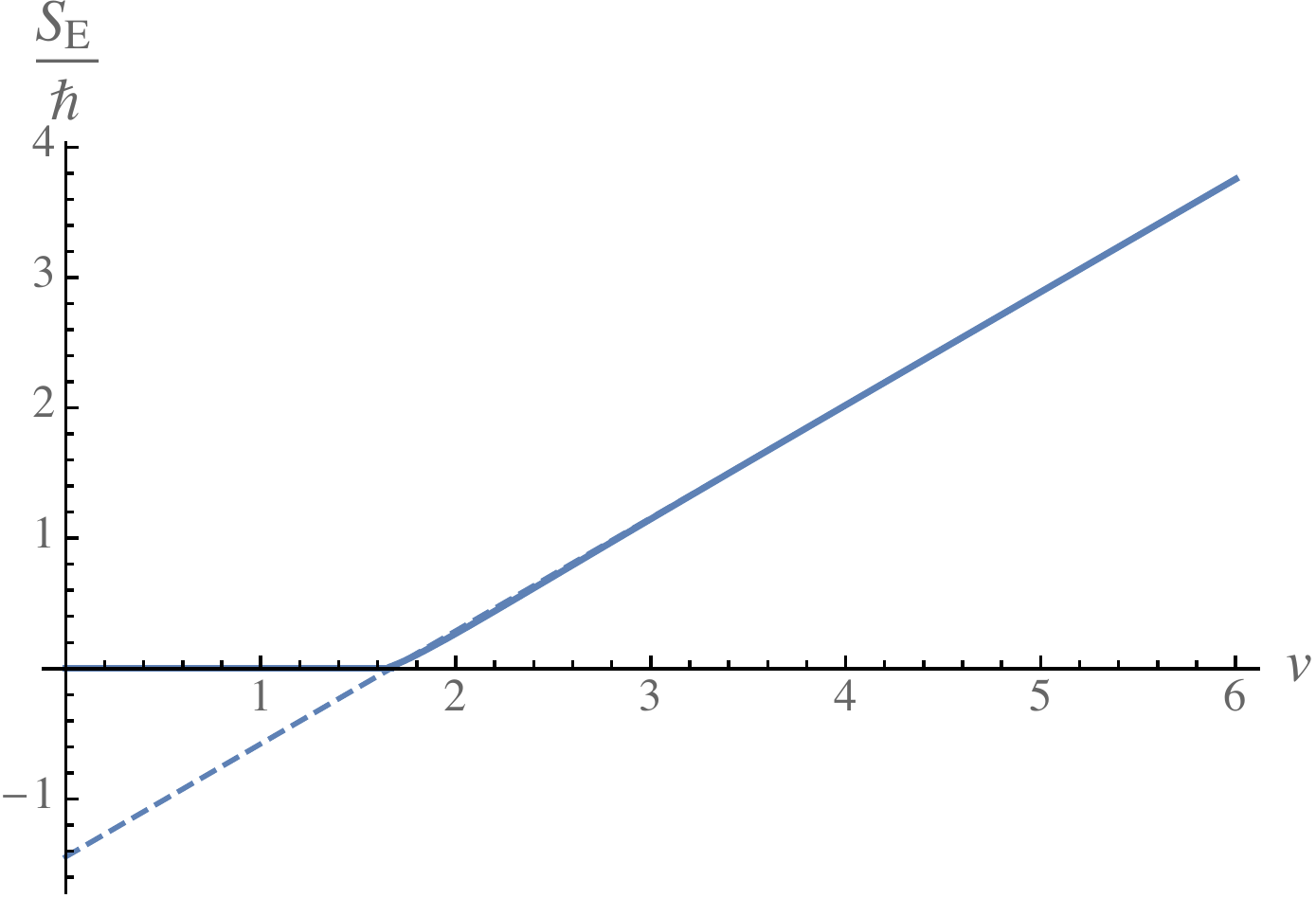}\caption{ The logarithm of the decay rate $\Gamma$ (times by $ -\frac{1}{2}$) versus the linear approximation in Eq.\eqref{eq:linearapprox} (shown as the dashed line) for $\gamma = 1$. The barrier disappears completely at $v=5/3$. \label{fig:nearlylinear}}
\end{figure}

How can we test this decay rate in a quantum annealer directly? The
assumption we will make is that the transverse field component of
the annealer induces an effective $\dot{\phi}^{2}$ term into any
field theory we encode on it, with some unknown coefficient. Therefore our method will be to construct on the annealer
a potential $U$ as given in Eq.(\ref{eq:potential}) and, by observing
its decay rates, test to see if the annealer has indeed turned it into
a $d=1$ QFT. The object of interest is therefore the exponent in
the decay rate:
\begin{align}
\hbar^{-1}S_{E} & =\int_{\eta_{+}}^{\eta e}\sqrt{\frac{2m(U-E_0)}{\hbar^{2}}}\,d\eta \\
 &\approx~  \gamma^{-\frac{1}{2}}\, \int^{\phi_e}_{\phi_+} \sqrt{\frac{3}{4}\tanh^{2}\phi-\text{sech}^2(\phi-v)}\,d\phi\,,\nonumber
\end{align}
where we have set $c=1$. Obviously this integral becomes linear in $v$ at large values, but a second advantage of the P\"oschl-Teller potential barrier  is that it remains so to a very good approximation, even for values of $v$ of order one, as shown in Fig.\ref{fig:nearlylinear}:
\begin{equation}
\label{eq:linearapprox}
\log\Gamma\,\approx\, -2\hbar^{-1}S_{E}\,\approx\,   \sqrt{\frac{3 }{\gamma}}  \left( \frac{5}{3} - v\right) \,.
\end{equation}
Thus we expect exponential decay
with an exponent falling linearly with $v$. Crucially this behaviour is qualitatively different
from thermal tunneling which has little dependence on the barrier
width $v$. For that one would instead expect to recover the Arrhenius equation, with
$\Gamma\sim e^{-E_{a}/kT}$, where $E_{a}$ is the activation energy
\footnote{This can be seen using the same techniques   \cite{Linde:1981zj}, but now the finite-temperature
field theory is genuinely Euclidean, with compactified time having
periodicity given by the temperature, namely $t_{E}=1/kT$. The instanton
has to satisfy the periodicity condition, and the time coordinate
is rescaled accordingly with $\beta=1/kT$. For high temperatures
there is effectively no room for derivative terms in the short interval
$\beta$, and we instead find
\begin{equation}
\Delta_{E}\sim\int_{\phi_{i}}^{\phi_{f}}\mathcal{D}\phi\,e^{-\beta\int_{0}^{1}dt(U-E_0)}\sim e^{-\beta E_{a}},
\end{equation}
with $E_{a}$ being the activation energy to reach the top of the
barrier. }.

\section{Implementation on a quantum annealer}

Let us now put together the components required to perform such a study. As mentioned
our goals are to encode the field theory potential $U(\phi)$ on the
annealer, then put the system into the approximate ground state of
a stable minimum, and add instability by adjusting the coupling
$k$ in Eq.(\ref{fig:PT-potential}). 

The method for encoding field theory was discussed in \cite{Abel:2020ebj}. In short
we begin with the effective Hamiltonian of the annealer, which is
a generalised Ising model of the form 
\begin{align}
\mathcal{H}_{QA}\,=&\,A(s)\left(\sum_{ij}\hat J_{ij}\sigma_{i}^{Z}\sigma_{j}^{Z}+C(t)\sum_{i}\hat h_{i}\sigma_{i}^{Z}\right) \nonumber\\ &+B(s)\sum_{i}\sigma_{i}^{X},
\end{align}
where $i,j$ label the qubits, $\sigma_{i}^{Z}$ are the $z-$spin
Pauli matrices, and $\sigma_{i}^{X}$ are the transverse field components,
while the couplings $\hat h_{i}$ and $\hat J_{ij}$ between the qubits are
set and kept constant. 

The reason these symbols are hatted is that they are not in general the ones $ h_{i}$ and $ J_{ij}$  that 
are input by the user. The annealer autoscales the latter until the largest absolute value of the couplings $h_i$ (resp. $J_{ij}$) is two (resp. one). That 
is 
\begin{equation}
\label{eq:rescale}
\hat h_{i} \, =\, \frac{h_{i}}{{\rm max}\{ |h_i|/2, |J_{ij}|\}} ~;~~
\hat J_{ij}\, =\, \frac{J_{ij}}{{\rm max}\{ |h_i|/2, |J_{ij}|\}} \, .
\end{equation}
In our study we will keep all the couplings sufficiently small that autoscaling is avoided (it is possible to extend the ranges of couplings but we will not do this here).

The parameter $s(t)$ (with $t$ being time)
is a user-defined control-parameter that can be adjusted during the
anneal, while $A,B$ describe the resulting change in the quantum
characteristics of the annealer, and $C(t)$ is another user-defined
parameter called the $h$-gain. To perform the more standard task
of finding a global optimisation, one would encode the problem to
be solved in the ``classical'' Ising model represented by the $A$-terms,
and then adjust the relative parameters $A,B$ in order to perform
an anneal from a highly quantum system to a classical one that has
$B=0$. For our purposes we will instead be probing the quantum properties
of the system when $B\neq0$. 

Scalar field values can be represented with the ``domain-wall encoding''
introduced in \cite{Chancellor19b}. That is we first add the Ising chain Hamiltonian:
defining the total number of qubits we use as $N$ (where $N$ should
be large), this is given by

\begin{align}
J_{ij}^{(\text{chain})} & =-\frac{\Lambda}{2}\,\left(\begin{array}{cccccc}
0 & 1\\
1 & 0 & 1\\
 & 1 & 0\\
 &  &  & \ddots\\
 &  &  &  & 0 & 1\\
 &  &  &  & 1 & 0
\end{array}\right),\nonumber \\
h^{(\text{chain})} & =\Lambda'\,\,(1,0,0\dots,0,-1),\label{eq:Jhchain}
\end{align}
where $\Lambda,\Lambda'$ are parameters that are somewhat larger
than the largest energy scale in the problem. (For the best performance
they should not be very much larger.) The coupling $h^{\text{(chain)}}$
forces the system to have spin $\sigma_{1}^{Z}=-1$ at one end, and
$\sigma_{N}^{Z}=+1$ at the other, while $J^{\text{(chain)}}$ forces
it to have as few spin-flips as possible. The result is a single ``frustrated''
position (the so-called domain wall) where the spin flips from negative
to positive. This position, $r$ say, encodes the value of the scalar
field as 
\begin{equation}
\phi\,\,=\,\,\phi_{0}+\xi r\,=\,\phi_{0}+\frac{\xi}{2}\sum_{i=1}^{N}(1-\sigma_{i}),
\end{equation}
where $\phi_{0}$ is a fiducial minimum value, while the second term
gives $r$ contributions of $\xi$ from the negative $\sigma_{i}^{Z}$
up to the domain wall position. It is then straightforward to see
that one can encode a potential term $U_{1}(\phi)$ in the $h_{i}$
couplings by adding
\begin{equation}
h_{j}^{\text{(QFT)}}=-\frac{\xi}{2}U_{1}'(\phi_{0}+\xi j).
\end{equation}
For our purposes, such a term cannot represent the whole of $U$ in Eq.(\ref{fig:PT-potential})
however, because we need to divide the potential into two pieces in
order to have the ability to turn on the metastable component. This
functionality is provided by the $h$-gain parameter $C(t)$, so the
entire potential is encoded as
\begin{equation}
U\,=\,U_{0}+U_{1}\,,
\end{equation}
where 
\begin{equation}
U_{0}\,=\,\frac{3}{4}\tanh^{2}\phi,\,\,\,;\,\,\,\,U_{1}\,=\,-{k(t)}\text{sech}^{2}\left(\phi-v\right),
\end{equation}
where $U_{0}$ remains to be encoded
in $J$. This allows us first to allow the system to settle in the
minimum around $\phi=0$, and then to adjust $C$ during the anneal
to turn on the potential $U_{1}$, and induce tunnelling. The encoding of $U_{0}$ into $J$ 
can be done by adding the couplings
\begin{equation}
J_{ij}^{\text{(QFT)}}=\frac{1}{4}U_{0}(\phi_{0}+\xi j)\left(2\delta_{ij}-\delta_{i(j-1)}-\delta_{(i-1)j}\right),
\end{equation}
where $\delta_{ij}$ is the Kronecker-$\delta$. These $J$ terms
contribute zero to the Hamiltonian except at the location of the domain
wall, where $(2\sigma_{k}^{Z}\sigma_{k}^{Z}-\sigma_{k}^{Z}\sigma_{k+1}^{Z}-\sigma_{k+1}^{Z}\sigma_{k}^{Z})=4$,
yielding a contribution $U_{0}(\phi)$ at that point. 

Note that $h^{(\text{chain})}$ is also scaled down when $C(t)$ is
small, so with this simple encoding we cannot set $C=0$. However
we do not need to initially turn off $U_{1}$ entirely, but just need
to reduce it so that tunnelling is not possible. A more precise encoding
that allows one to turn off $U_{1}$ entirely
is to share $U_{1}$ between $J$ and $h$ such that the initial value
of $C$ makes them cancel exactly. That is
\begin{align}
\label{eq:divide}
J_{ij}^{(U)} =&   \frac{1}{4}\left[U_{0}(x_{0}+\xi j)-\frac{C_{0}}{1-C_{0}}U_{1}(x_{0}+\xi j)\right] \nonumber \\
                   &  \left(2\delta_{ij}-\delta_{i(j-1)}-\delta_{(i-1)j}\right),\nonumber \\
h_{j}^{(U)} = & -\frac{\xi}{2}\frac{1}{1-C_{0}}U_{1}'(x_{0}+\xi j),
\end{align}
where the choice of parameters $C(0)=C_{0}$ and $C(t_{f})=1$ gives
the desired behaviour. We shall use this later on but for the moment we will stay with the simpler assignment of potentials.

\begin{figure}
\makebox[0pt]{
\centering{}\includegraphics[scale=0.48]{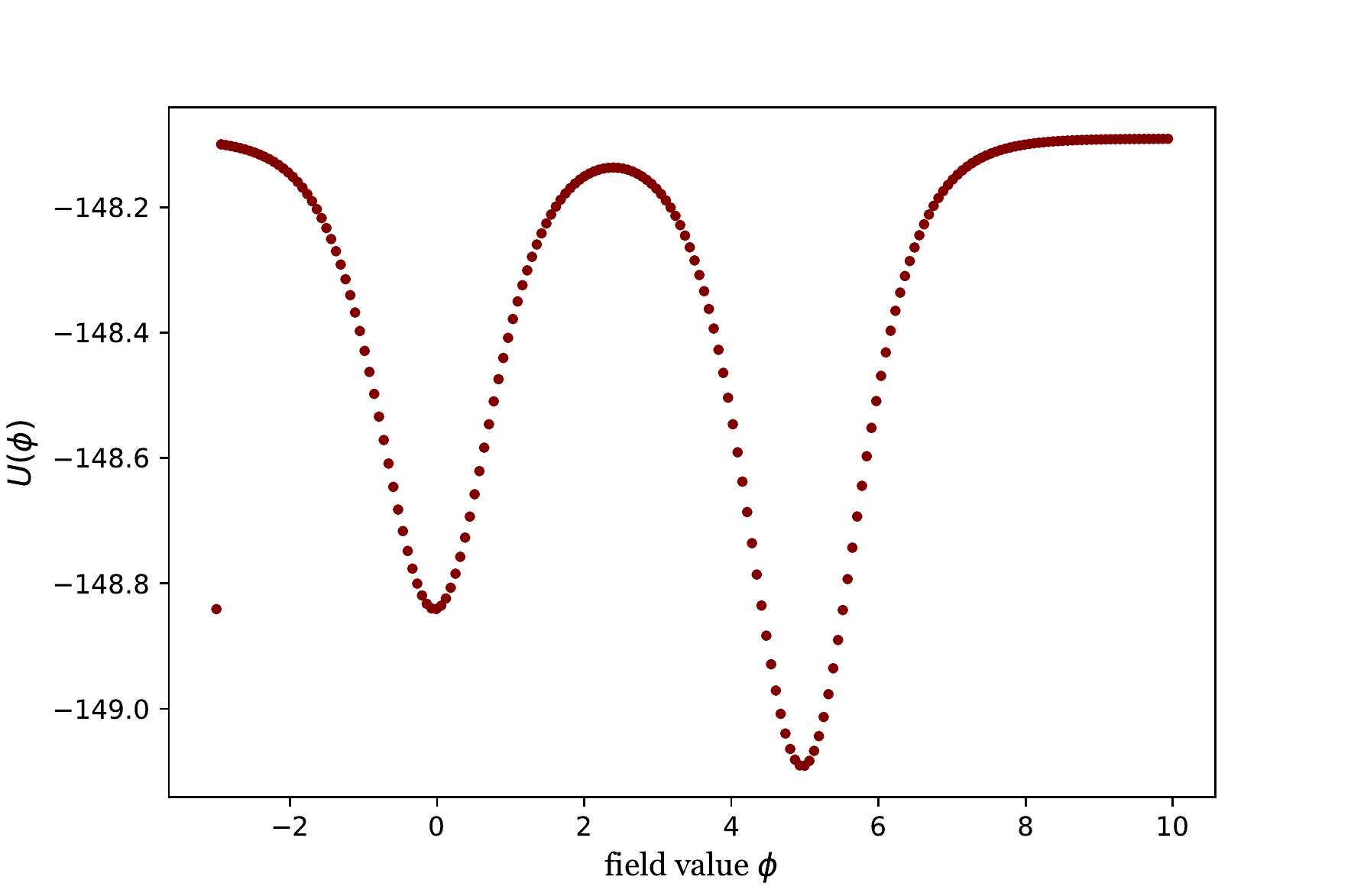}}\caption{The potential as seen by the Ising model on the annealer, where we choose $N=200$ qubits, and parameters $k=1$ and $v=5$, c.f. the actual potential in Fig.\ref{fig:PT-potential}. Note the large negative
 overall energy off-set due to the field theory encoding, and the ``dropped qubit'' at $\phi=\phi_0$.  \label{fig:isingview}}
\end{figure}

This completes the encoding of the field theory potential.
To verify that it is working as desired, we show the resulting potential in Fig.\ref{fig:isingview}.
For this and the remainder of the work we shall take $N=200$ as a reasonable compromise between 
accuracy and efficiency on the annealer. 
As expected there are two unavoidable features of the Ising potential compared to the original 
one, both caused by the Ising chain encoding of the field theory: first the negative rewards in
$J^{\rm chain}$ cause an off-set of order $-N \Lambda$; second the rewards in $h^{\rm chain}$
in Eq.\eqref{eq:Jhchain} imply ``dropped qubits'' at the first and last positions (the one at the 
last position is off the scale). Neither of these should affect the tunnelling rate.

\begin{figure}
\centering{}\includegraphics[scale=0.27]{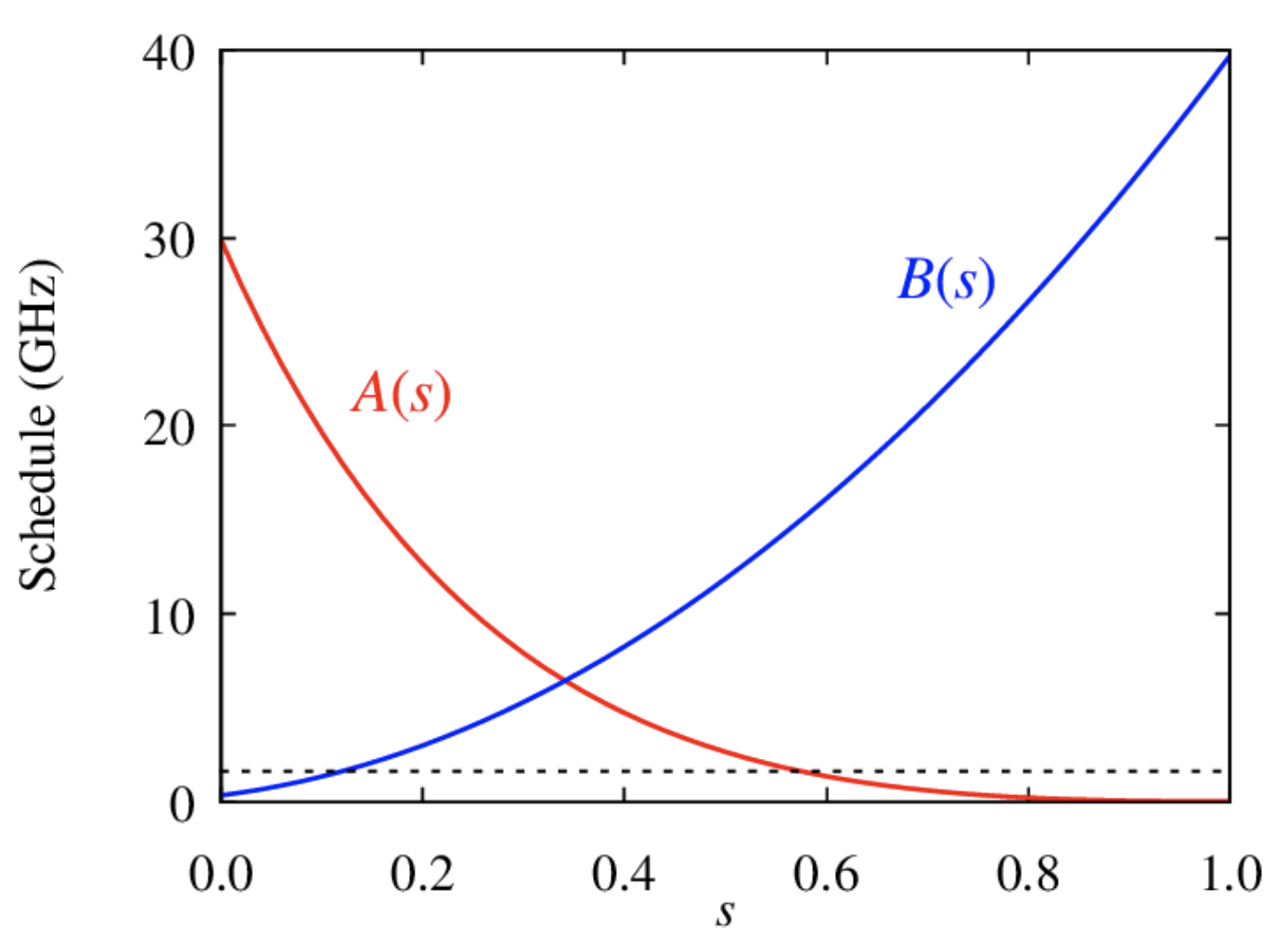}\caption{Anneal schedule parameters. The thermal contribution is shown as a
solid line, while $A$ and $B$ are the coefficients scaling the classical
Ising and transverse field contributions respectively  \label{fig:Anneal-schedule-parameters}}
\end{figure}

\begin{figure}
\centering{}\includegraphics[scale=0.5]{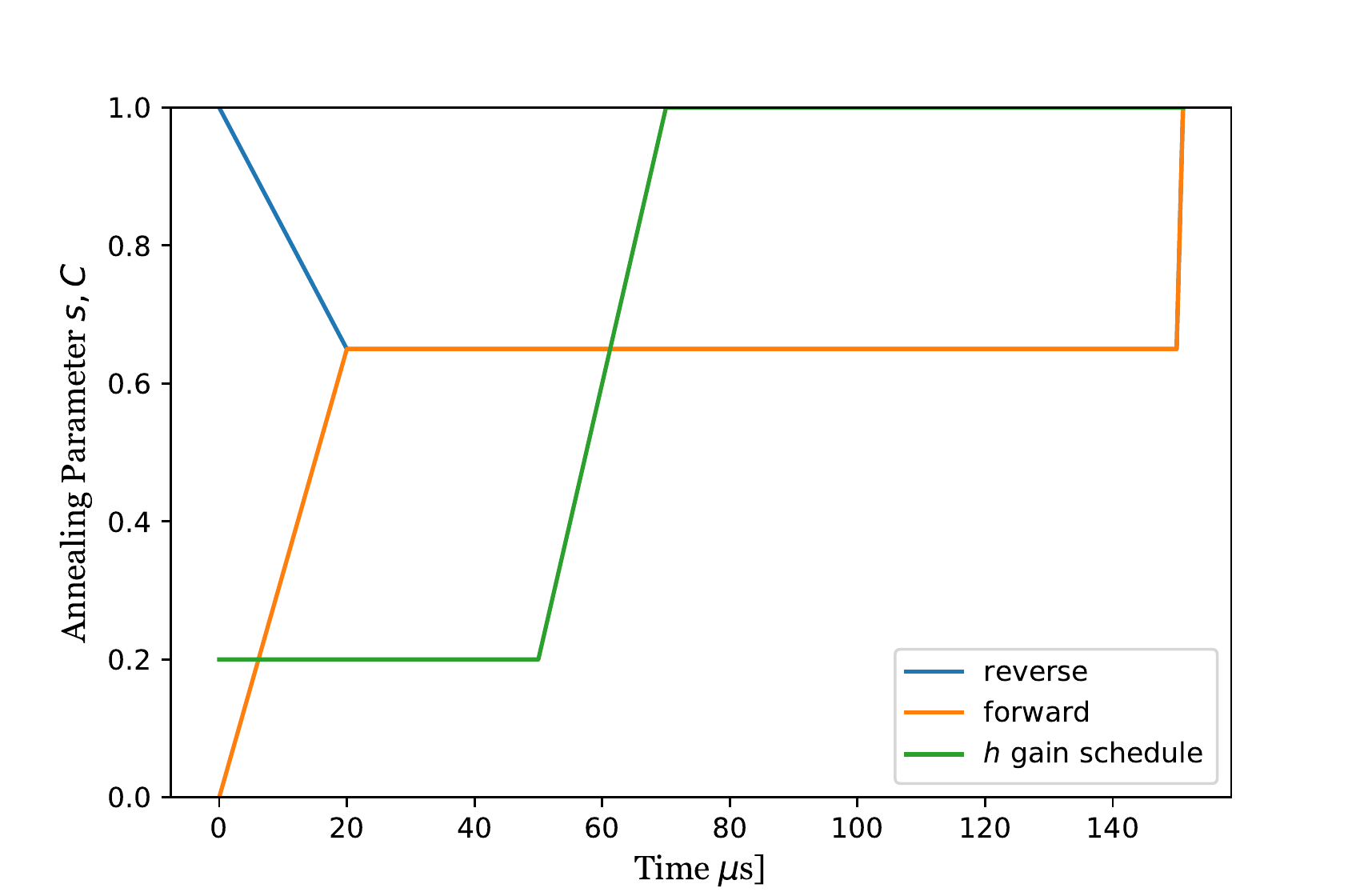}\caption{Typical reverse anneal schedule. The anneal parameter $s$ increases
the transverse field, and there is an initial period of stabilisation
in the minimum at the origin. The $h$-gain parameter is then turned
on to introduce metastability and induce tunnelling.\label{fig:Anneal-schedule-parameters-1}}
\end{figure}

Let us now turn to the configuration of the anneal itself. As mentioned,
the coefficients $A$ and $B$ describe how ``quantum'' the system
is, and are best visualised with the plot in Fig.\ref{fig:Anneal-schedule-parameters}.
When $s=0$ the system is maximally quantum, and when $s=1$ the system
has arrived at the pure classically Ising-encoded problem. A ``forward
anneal'' schedule would take $s(t_{i})=0$ and $s(t_{f})=1$,
beginning with a rapidly tunnelling system, and ending up at a system
that solves the optimisation problem of interest. A ``reverse anneal''
schedule gains initial classical control with $s=1$. Then we turn
on the quantum mechanics so that we send $s$ to some finite value
$s_{q}$ for some time-interval, before returning to the classical system. 
This latter option is the one we choose, as it allows us to fix the system in the 
false vacuum, and then count the number of times it tunnels when it is sent for a given 
period to $s_q$. It is shown beginning as the blue line in Fig.\ref{fig:Anneal-schedule-parameters-1}, returning to 
$s=1$ on the orange line.

Note that the value of $s_{q}$, i.e. the regime where we induce quantum
mechanical behaviour, is much larger ($s_{q}=0.7$ in the figure)
than would normally be the case. In fact Fig.\ref{eq:groundstate}
makes it clear that we will choose it to be where quantum mechanics
is just turning on, in order to have relatively slow tunnelling and maintain
good control. 

During the anneal we will choose an \emph{h}-gain schedule, $C(t)$,
which varies between $C_{0}<1$ and $1$, as indicated by the green
line in Fig.\ref{fig:Anneal-schedule-parameters-1}. For an initial
period the $h$-gain begins at a small enough value such that the
second minimum induced by $U_{1}$ is higher than that at the origin
$U_{0}$. During this initial relaxation and dissipation period the system is unable to tunnel, so ultimately it
is expected to reach the ground state of $U_{0}$ given
by Eq.(\ref{eq:groundstate}). Once it is in a stable bound state we can
adjust $C(t)$ to send the coupling $k\rightarrow 1$, and turn on tunnelling for the rest of the anneal.
This configuration, in which we first allow the system to
settle, is forced on us by the quantum properties
of the annealer. Indeed if we were to start the system at the bottom
of the metastable minimum at the origin and then simply turn on the
transverse field, it would tunnel very rapidly. This is because in
a reverse anneal the classical starting point is a pre-defined set of $\sigma_{i}^{Z}$'s.
This implies that the initial wavefunction $\psi(\phi)$ is a position eigenstate (it is essentially
a Dirac $\delta$-function), containing superpositions of all energy
eigenstates. 

It is worth 
 mentioning several moves that are required to improve performance. For all our results we will 
 using a minor-embedding on the Dwave annealer QPU, due to its limited connectivity, with $N=200$ qubits
in our effective Ising model (but obviously with more on the physical
machine due to the embedding). Performance is improved by splitting the large number of reads into smaller groups (of say 100) 
in order to reduce biasing from each embedding. The states are re-initialised at the bottom of the 
 false vacuum in a classical state at the beginning of each read. 
As mentioned one also has to be careful to
set the Ising chain parameters, namely $\Lambda,\Lambda'$, to be
not much larger than the largest energy scale in the problem. This
is because as mentioned we wish to avoid the annealer autoscaling the couplings to $\hat h,\hat J$ as in \eqref{eq:rescale}. 
After such scaling, Ising chain parameters that were very large, would imply couplings in the 
physical potential that were very small. The effect of autoscaling is actually an additional motivation for our favouring of P\"oschl-Teller potentials,
because they go to a constant at large field values and different $
\phi$ intervals do not change the autoscaling: by contrast a
quartic potential would grow rapidly at large field values\footnote{It is also worth mentioning that the D-Wave annealer does 
provide the possibility of turning off auto-scaling (by setting $ {\tt auto_{-}scale = False}$) but the performance is reduced
unless the couplings are tuned precisely anyway.}. Conversely if the Ising chain parameters
are too small then the Ising chain breaks and we no longer have a
faithful representation of the field value. Such ``wall-breaks''
happen a few percent of the time and can never be eliminated entirely. Those results are simply discarded.
Additionally the minor-embedding itself (which ties qubits together
in a similar fashion to the Ising chain embedding in $J$) may also fail. 
The parameters can usually be adjusted so that these ``chain-breaks'' happen rarely however.

\section{Results}

\subsection{Calibration on SHO ground states}

\begin{figure}
\centering{}\includegraphics[scale=0.54]{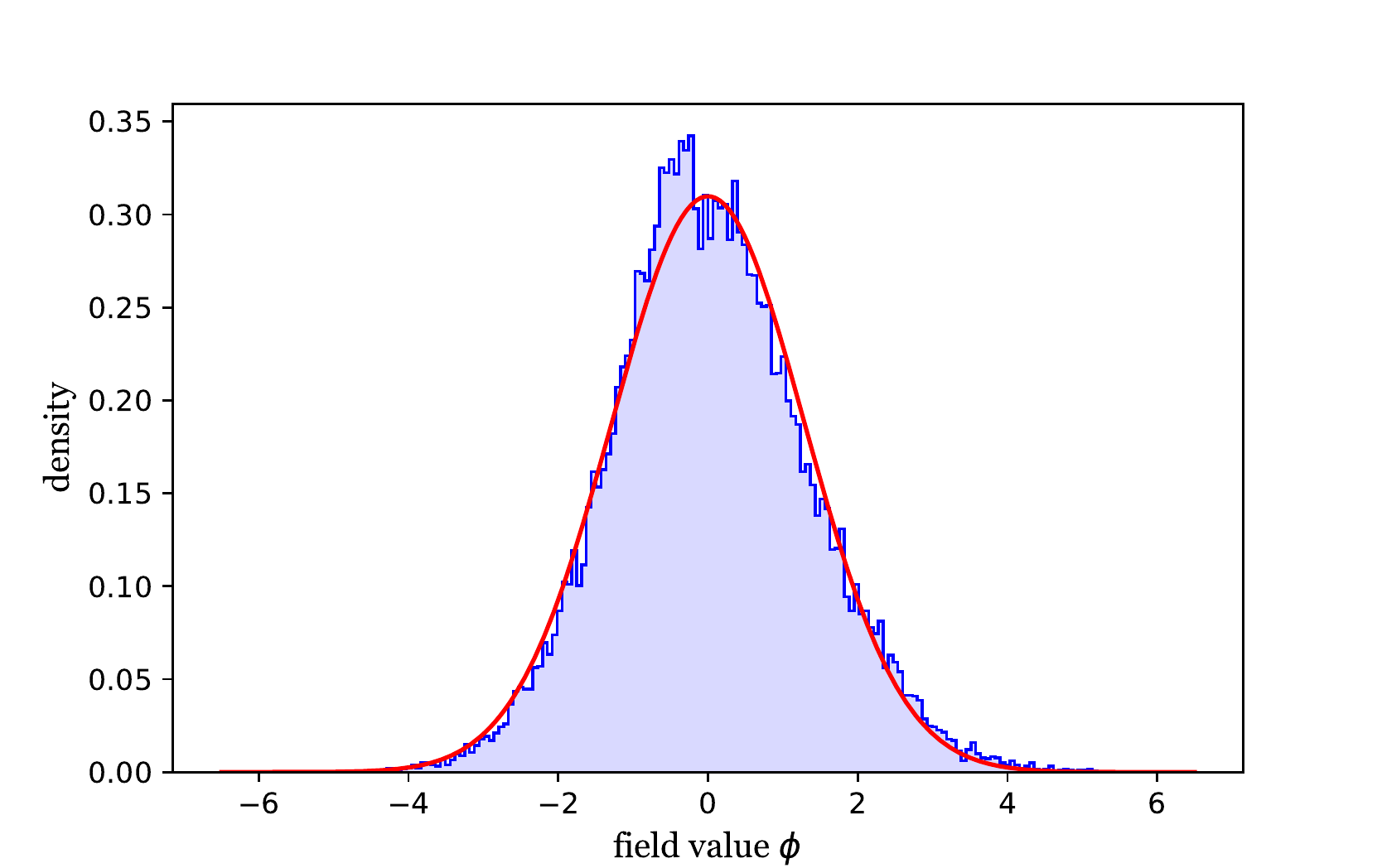}
\caption{The probability density of the SHO with $N=200$ and with $s_{q}=0.7$ after time $t=75\mu$s and with $\kappa=0.06$. The ground-states are measured with an interval $\Delta\phi = 13$. The probability density approximates the red line, which corresponds to $\gamma \equiv  \hbar^2/2m\eta_0^2 = 0.33$.\label{fig:groundstateab}}
\end{figure}

We now turn to the results, and discuss the various parameters and further interpretation 
as we proceed, beginning by studying the system with no tunnelling. That is we
keep $C(t)=C_{0}$ and set $v$ to be very large, in order to learn about the effective Planck's constant, more precisely the 
combination $\gamma=\hbar^2/2m\eta_0^2$. As mentioned this amounts to our calibration of the experiment, and to perform it in a systematic way, we will use the simple-harmonic-oscillator (SHO).
That is we take 
 \begin{equation}
U_0(\phi)\,=\, \frac{\kappa }{2}\phi^2\, .
\end{equation}
We show the result of 30K reads of the annealer with 
$\kappa=0.06$ in
 Fig.\ref{fig:groundstateab}, presented as binned probability density functions normalised to one. (In other words as $N\rightarrow \infty$ this curve would be $|\psi|^2$). Note that the value of $\kappa $ is chosen small enough to avoid autoscaling. For this run we hold the annealer at $s_{q}=0.7$ for 75 $\mu$s (plus 5 $\mu$s of ramp-up and 1 $\mu$s of ramp-down). 
 
By inspecting this and similar curves one gains some intuition about  the behaviour of this system. 
First, apart from some seemingly characteristic perturbation around the peak it clearly appears to have reached the Gaussian ground state,
which is of the form 
\begin{equation}
\label{eq:SHOGS}
|\psi|^2 \, =\, \frac{( \kappa/2\gamma)^{\frac{1}{4}}}{\pi^{\frac{1}{2}}} e^{ - \sqrt{ \kappa/2\gamma} \,\phi^2 }~,
\end{equation}
 so we can reasonably conclude that for this choice of parameters 75 $\mu$s is long enough for the required dissipation. 
 Note that the $\eta_0$ parameter cancels in the $\kappa/\gamma$ ratio.
 Secondly, this curve leads to an approximate estimation of $\gamma = 0.33$. Choosing different physical couplings appears to yield similar values of $\gamma$, so not only do the wave-functions have the correct shape but they also have the correct  functional dependence on $\kappa$.  By contrast the result for the inferred value of $\gamma$ {\it does} depend on the interval we choose for $\phi$. This is because different intervals with the same choice of $N=200$ imply different $\xi$, 
and not surprisingly this affects the mass density $m$ in the field theory.

We stress
that absolutely no dynamics was introduced by hand into the annealer,
and therefore this constitutes a genuine measurement of the ground
state wavefunction of a quantum mechanical system. 

It is also instructive to consider the fact that the annealer returns a
wave-function with different $\gamma$ depending on the value of $s_q$. 
When we choose $s_{q}$ we imbue the effective field
theory with a kinetic $\dot{\phi}^{2}$ term that has a certain value
of $\hbar^{2}/2m$ we do not know. The ground state has to adjust
to have the matching value of $\gamma$. Clearly as we let $s_{q}\rightarrow1$ the value
of $\hbar^{2}/2m$ in our effective theory must go
to zero because quantum effects turn off there. Accordingly the ground state wave-function becomes increasingly narrow
until in the classical limit it approaches a $\delta$-function, which in a reverse anneal is where it begins. In other
words the ``classical'' $\delta$-function position eigenstate is simply the ground
state wave function when there is no transverse field component.

\subsection{Tunnelling }

\begin{figure}
\centering{}\includegraphics[scale=0.6]{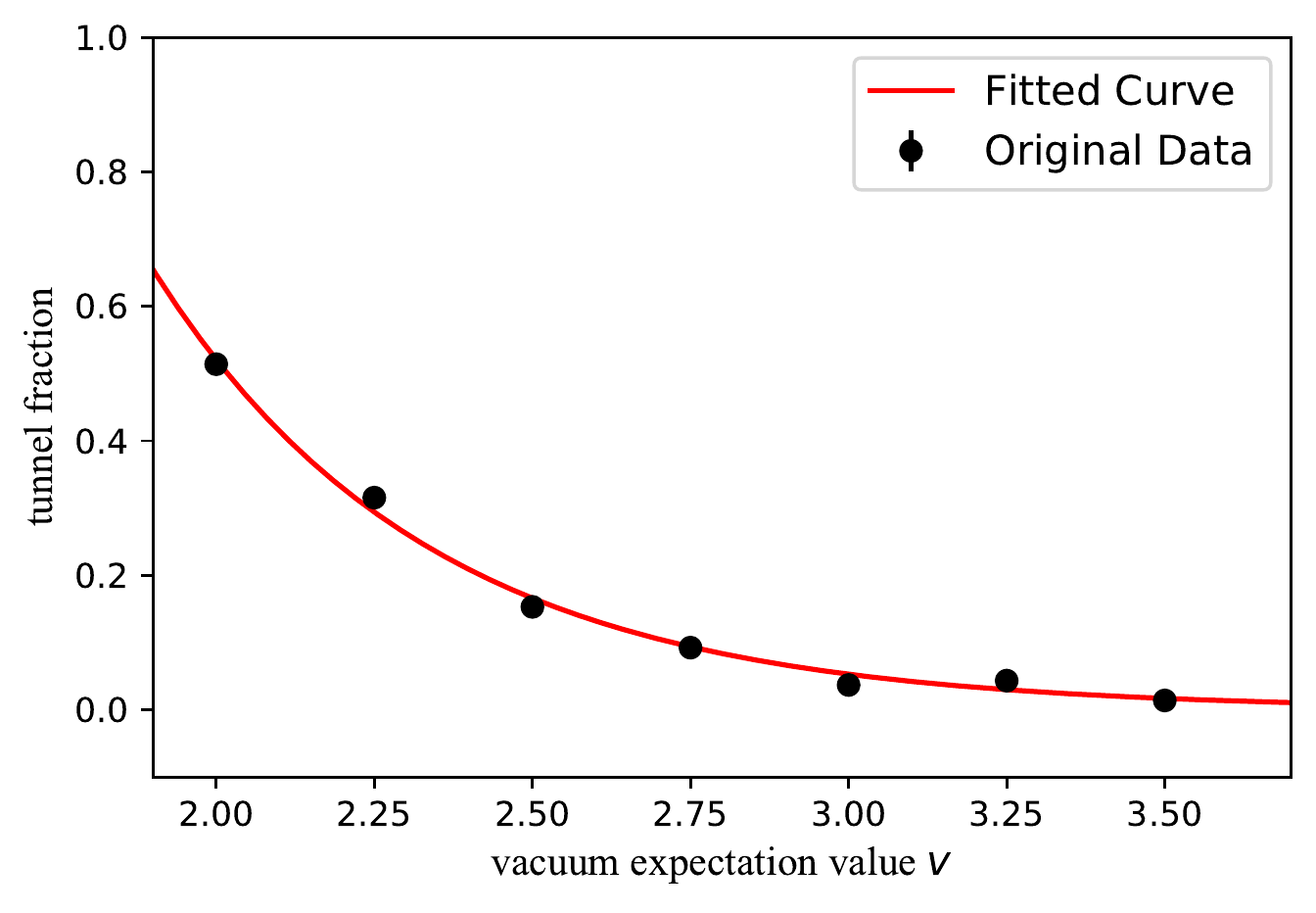}\caption{Best fit values for the tunnelling fraction $P(v)=a e^{-b v}$ for varying vacuum expectation values $v$, with tunnelling time $t_{\mathrm{tunnel}}=100\mu s$ are $a=50.5$ and $b=2.29$. \label{fig:fit}}
\end{figure}

We now turn to our double-well potential, and adjust the $h$-gain schedule so that, after setting, the second minimum appears and the system is able to tunnel into it for a period $t_{\text{tunnel}}$. One can perform the same exercise as for the SHO ground state. The result (now displayed as a probability distribution such that the 
sum of the bin-counts is normalised to unity) is shown in Fig.\ref{fig:tunnelled},
for the system when it is left for $50,100,150\mu$s in the presence of the second minimum, with $k=1$ in
the potential of Eq,(\ref{fig:PT-potential}), where we take $v=2.5$. The presence of tunnelling is clearly evident. 
Further evidence in support of this being genuine quantum tunnelling can be  found by studying the 
decay rates as a function of $v$. This is shown in Fig.\ref{fig:tunnelled} for several values of $v$ where the expected exponential 
suppression of the decay rate with increasing $v$ is apparent. This exponential behaviour can be fit to the approximation in \eqref{eq:linearapprox}, as 
in Fig.\ref{fig:fit}. For the measured value of $\gamma$ the theoretical expectation is $\log\Gamma = 3.0 \times (1.66 - v)$. The best fit value
(given by the red line in  Fig.\ref{fig:fit}) is $\log\Gamma = 2.29 \times (1.71 - v)$. Perhaps unsurprisingly, the overall parameter $\gamma$ remains one of the 
most difficult aspects to determine precisely given the limitations of the annealer for this study. Nevertheless the observed behaviour 
provides good support for the presence of quantum tunnelling.

\begin{figure}
\centering{}\includegraphics[scale=0.6]{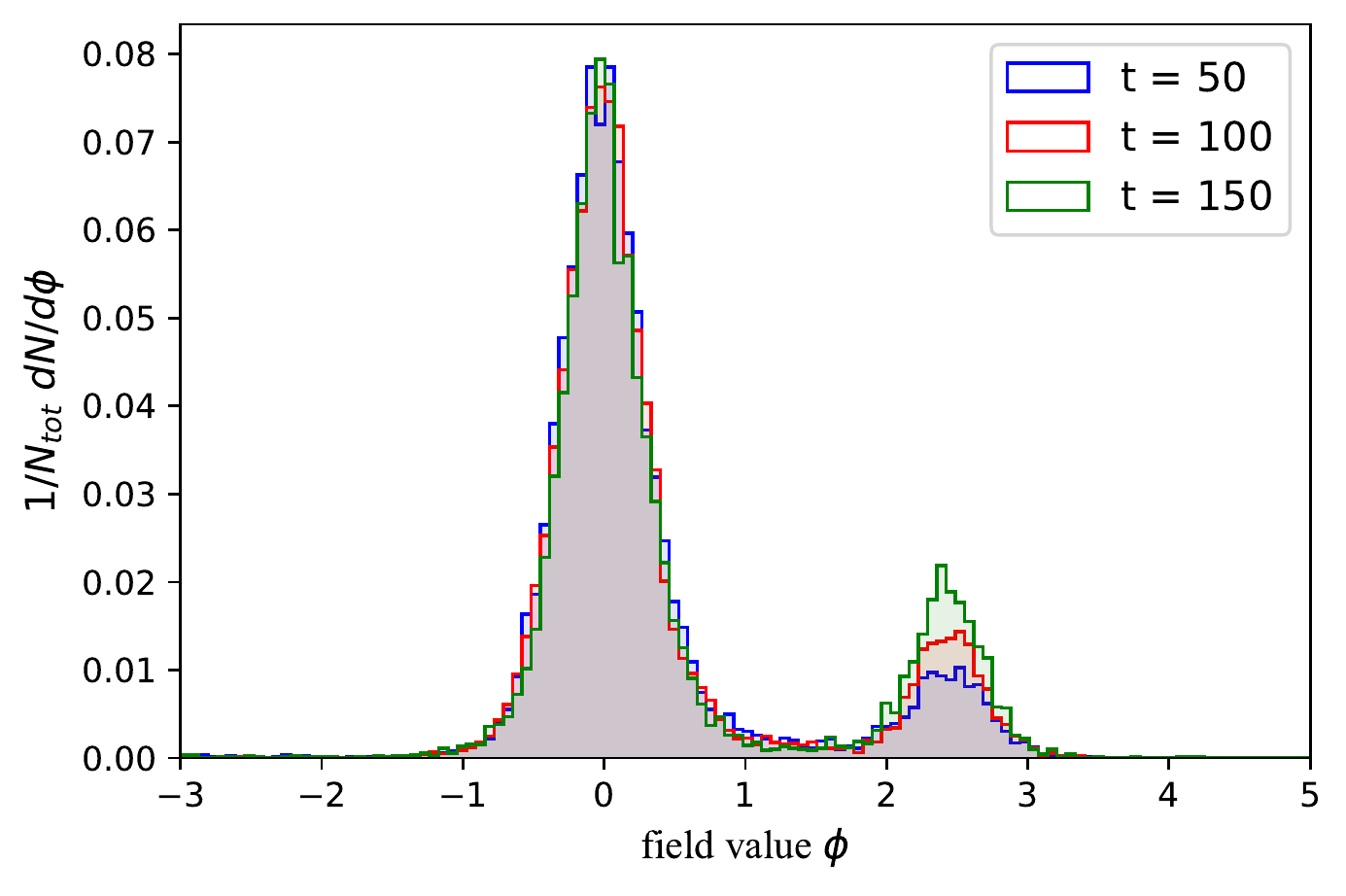}\caption{The probability distribution  with $v=2.5$, $s_{q}=0.7$ after $t_{\text{tunnel}}=50,100,150mu$s, where $N$ is the number of events\label{fig:tunnelled}.}
\end{figure}

\begin{figure}
\centering{}\includegraphics[scale=0.62]{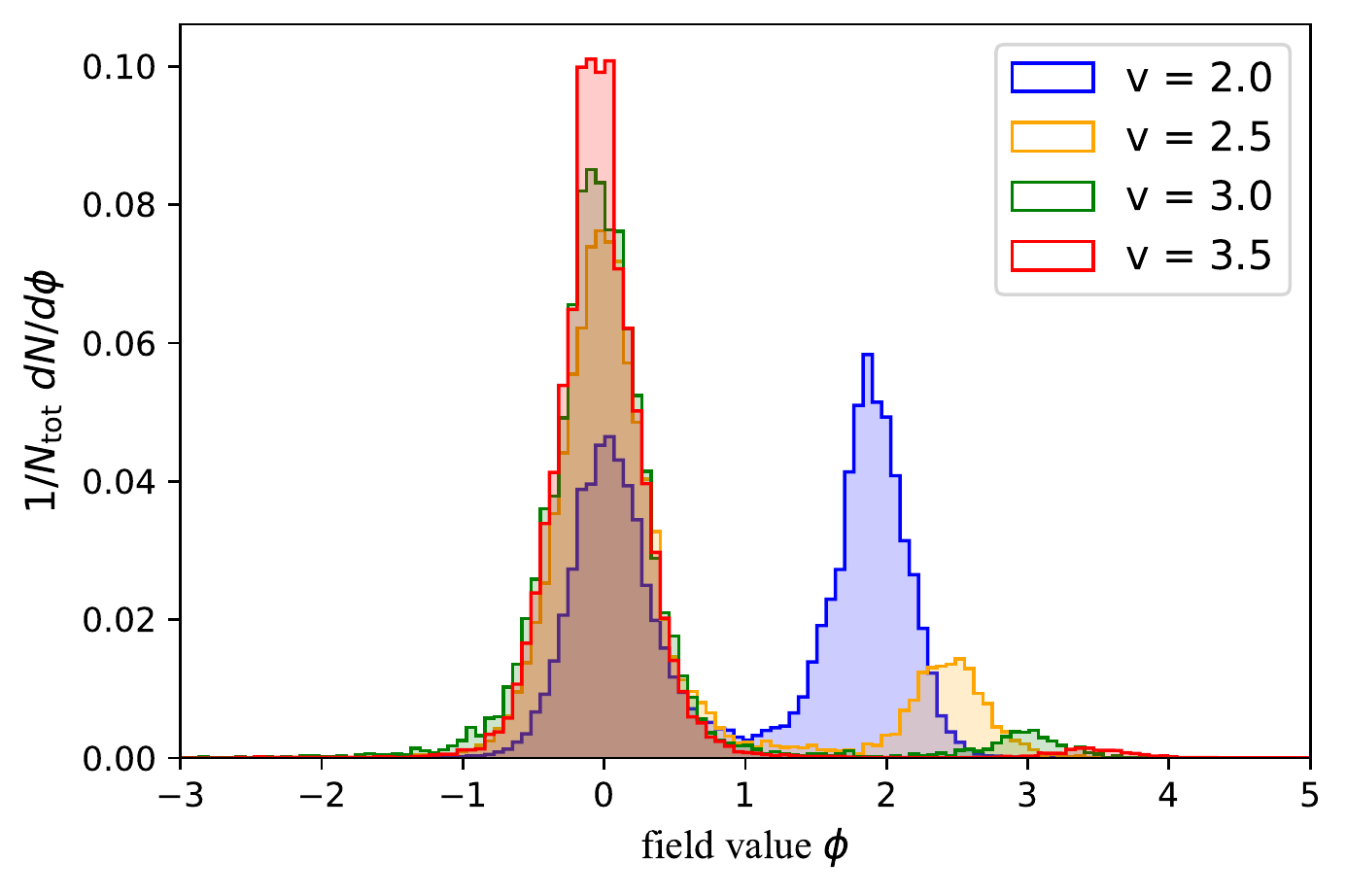}\caption{The transition probabilities for different $v$ with $s_{q}=0.7$ after $t_{\text{tunnel}}=100\mu s$.\label{fig:tunnelled}}
\end{figure}

\subsection{Quantum versus Thermal}
It is important to definitively exclude the possibility that what is being observed is thermal rather than quantum 
tunnelling. More precisely we wish to establish that the states are really tunnelling through the barrier rather than being thermally 
excited over the top, noting for example that an explanation for the drop-off with $v$ observed in the tunnelling rate above, could simply be due to the 
height of the barrier (and hence the activation energy $E_a$) increasing with $v$. 

\begin{figure}
\centering{}
\makebox[0pt]{
\includegraphics[scale=0.45]{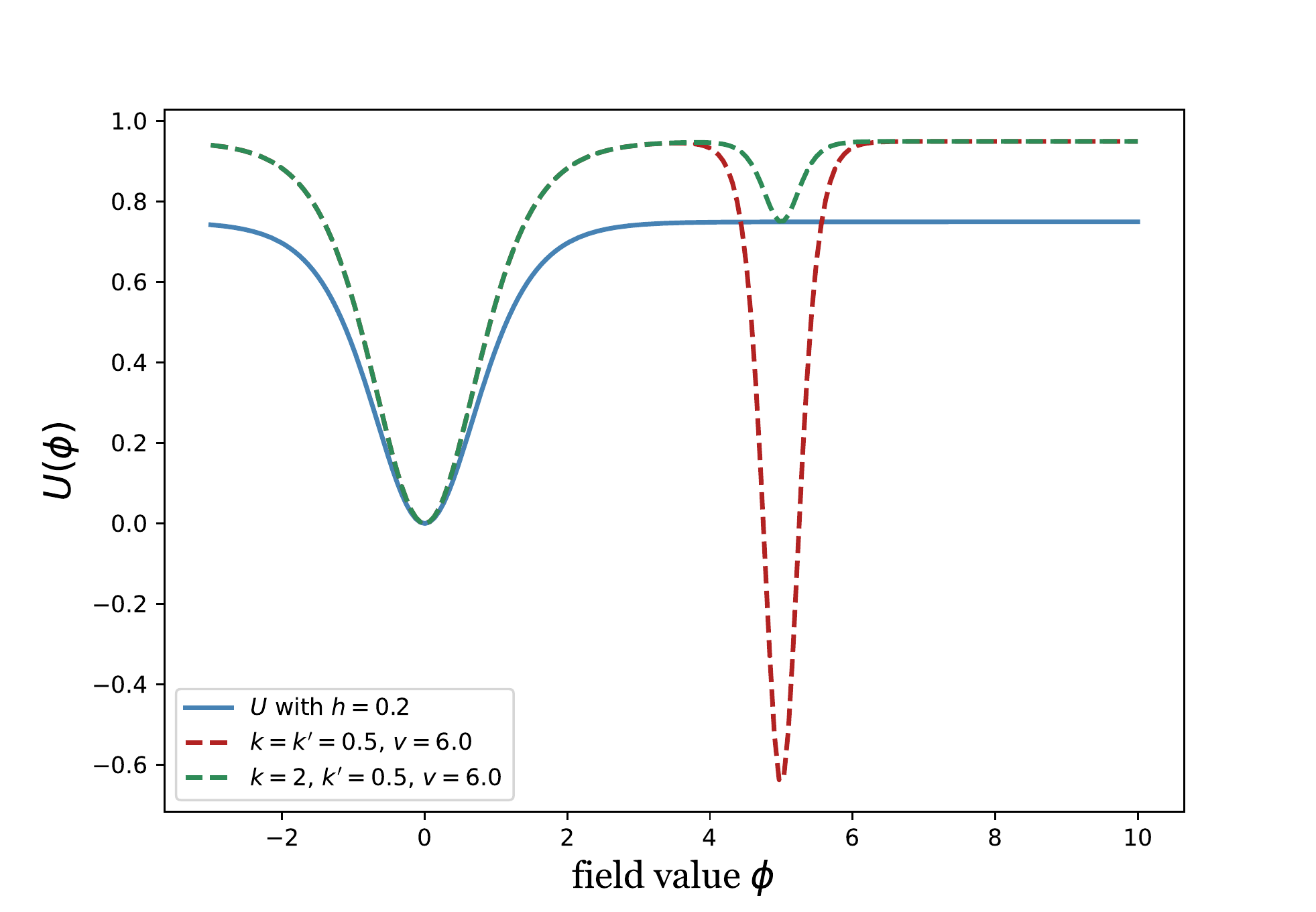}}
\caption{Minimally disturbing the initial state in order to test if the tunnelling exhibits quantum or thermal behaviour. The initial 
potential is a single well, and additional terms raise a barrier between it and a new well that is introduced with either a minimum at either exactly the same height as the 
original potential, or deeper than the original one. \label{fig:wall-potential}}
\end{figure}

In order to probe this particular question,  we will now examine a potential that provides a cleaner separation between quantum and thermal behaviour, as shown in Figure \ref{fig:wall-potential}. 
The potential is divided up more precisely than before, in the manner  described earlier, so that it is of the form  in \eqref{eq:divide} where we take $C_0=0.2$ as our initial $h$-gain parameter. 
In other words the terms in our new potential can be written  
\begin{align}
U_{0}&\,=\,  \frac{3}{4} \tanh^{2}\phi - C_0 \,U_{1}~, \nonumber \\
U_{1}&\,=\, k'\, \tanh^{2}\phi - k \, \text{sech}^{2} c(\phi-v)~,
\end{align}
with the potential at $t=0$ being the single P\"oschl-Teller well, shown as the solid blue line.
When $C(t)\rightarrow 1$, the first term in $U_1$ then raises the sides of the well by $(1-C_0) k'$, while the second term introduces a new well at $\phi=v$ of width $\sim 1/c$ and depth $(1-C_0)k$. 
We will take $c=3$ and $k'=1/2$. We then consider $k=k'$ or $k=2$. For this study we will also choose $s_q=0.65$ which gives more rapid tunnelling, allowing us choose values of $v$ that are in the flat region of the potential.

There are several reasons that this constitutes a clean separation of quantum and thermal behaviour. First it is notable from the study above that the 
bound state in which the system begins has a rather high energy. As such if we simply introduce a new minimum as we did earlier then it is likely that some components of the 
wave-function will be able to tunnel rapidly. The initial dip at $v$ that was present in our previous configuration would also be able to capture states during the dissipation phase. Neither of these two types of state could be very easily distinguished from ones that had thermally tunnelled. 

What do we expect the tunnelling behaviour to be in the potential above? In the situation where $k=k'$ no new minimum is introduced that would be quantum mechanically accessible to any component of the initial bound state. Therefore in principle we should not find any states in this minimum at all if the system is purely quantum, although in practice this will depend on there being no remaining continuous component in the spectrum at all. This is in contrast to the case where $k=2$ shown as the dashed red line in Fig. \ref{fig:wall-potential}, where the standard quantum tunnelling should take place. Moreover according to \eqref{eq:linearapprox} the observed tunnelling rate into this minimum should again drop-off with increasing $v$, even if we consider values of $v$ in the region where barrier height is constant.

Let us contrast this behaviour with what one would expect for a thermally activated system. In this case there would be little distinction between the $k=1/2$  and $k=2$ cases. Once thermal effects are large enough to excite states over the barrier, roughly similar proportions would be captured by the new minimum at $\phi=v$. How much remains trapped there depends somewhat on the temperature and whether the transitions are in equilibrium. Calling the minima at $0$ and $v$, $A$ and $B$ respectively, and the height of the barrier $E_a$, ultimately such a system would attempt to reach an equilibrium where the transition rates are the same in both directions, i.e. $\frac{N_A}{N_B} = e^{E_a /k_BT} e^{- (E_a-E_B))/k_BT} = e^{E_B/k_BT}$.  If the system were fully in equilibrium then the ratio of the numbers of states found in the new minima would be independent of the height of the barrier, and of order $e^{ (E_{B_1}-E_{B_2})/ k_BT}$, where $1,2$ labels the choice $k=k'$ or $2$ respectively.  However the difference in energies  $(E_{B_1}-E_{B_2})$ is of the same order as the activation energy $E_a$ itself. Therefore a significant thermal tunnelling would result in similar numbers of states in the new minima. And the $k=2$ and $k=k'$ cases become only more similar if the transitions begin to fall out of equilibrium, as the rate of tunnelling in either direction would become very low: the number count in the new minimum would then simply depend on how many states had fallen into its domain of attraction, and this would be virtually independent of the depth. Finally the tunnelling rate should not depend on $v$ in this potential if it proceeds by thermal activation: any thermally activated state would be equally likely to fall into the new minimum regardless of $v$.

\begin{figure}
\makebox[0pt]{
\centering{}\includegraphics[scale=0.57]{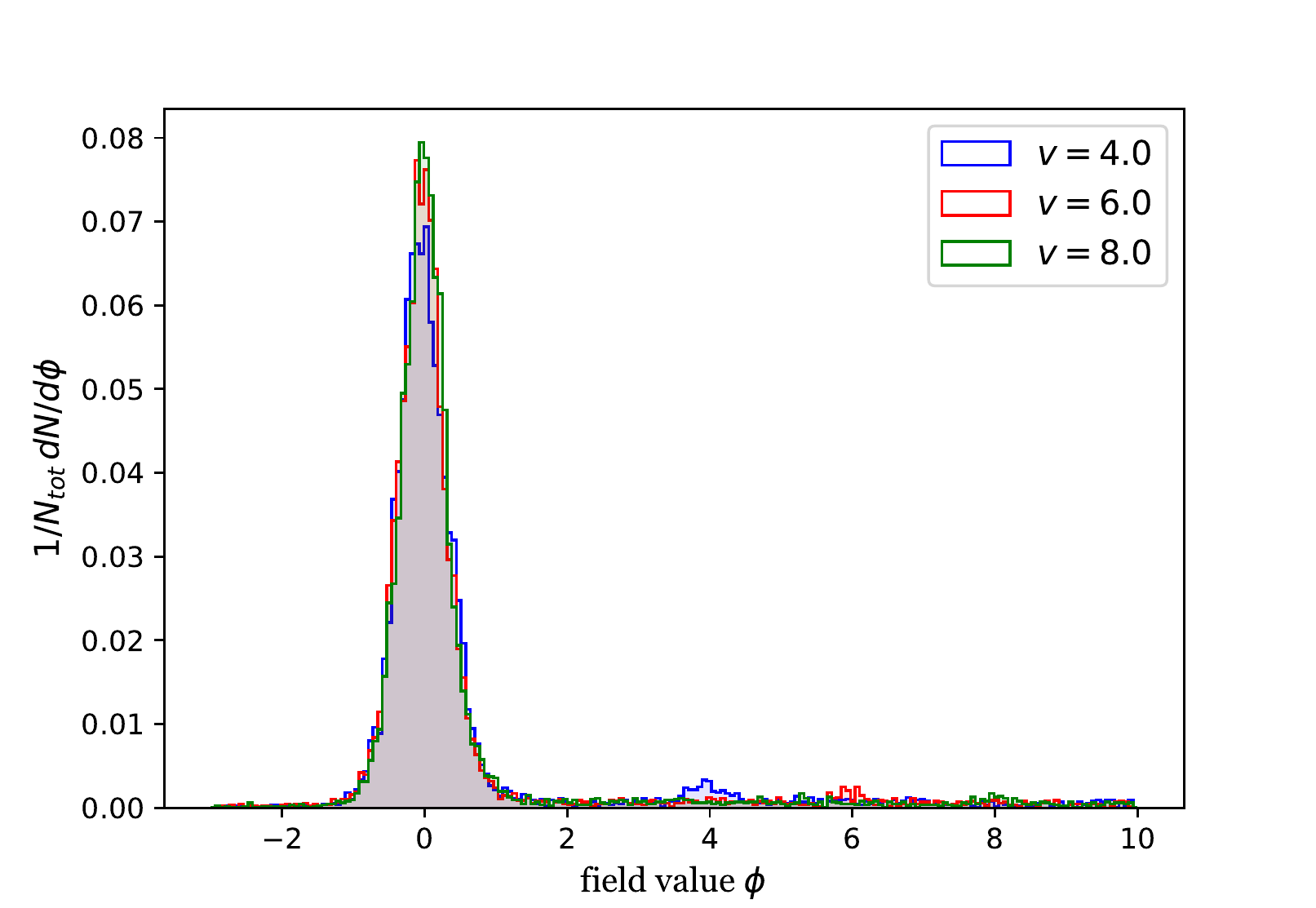}}\caption{The transition probabilities into the raised minimum of Fig.\ref{fig:wall-potential} for $v=4$ with $s_{q}=0.65$ after $t_{\text{tunnel}}=100\mu$s.\label{fig:QuantumVThermal2}}
\end{figure}

\begin{figure}
\makebox[0pt]{
\centering{}\includegraphics[scale=0.57]{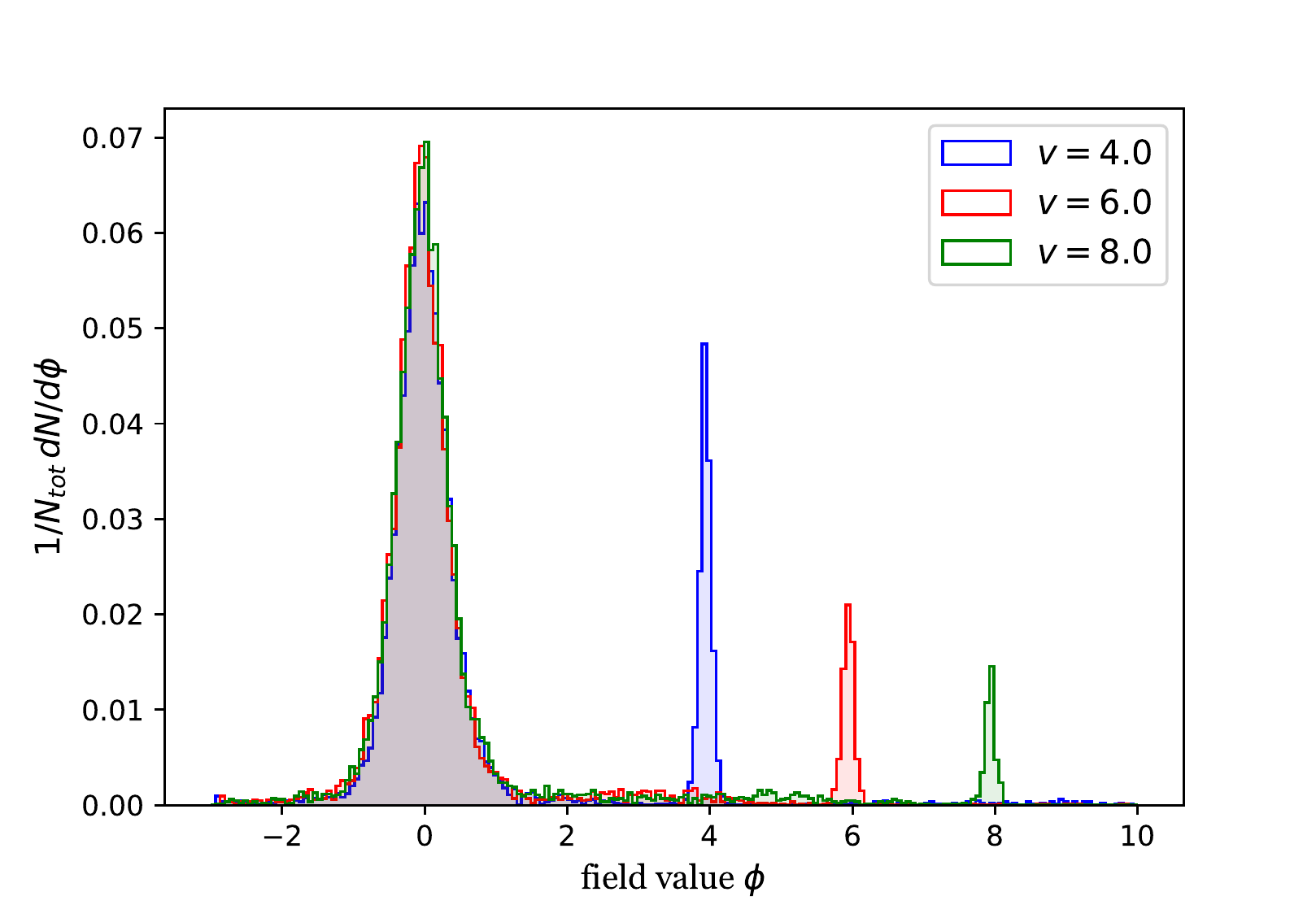}}\caption{The transition probabilities for different $v$ in the presence of the deep minima of Fig.\ref{fig:wall-potential}, with $s_{q}=0.65$ after $t_{\text{tunnel}}=100\mu$s.\label{fig:QuantumVThermal1}}
\end{figure}

Results from the two cases $k=k'$ and $k=2$ are shown in Figures \ref{fig:QuantumVThermal2} and \ref{fig:QuantumVThermal1} respectively. The former shows the expected quantum tunnelling behaviour with a rapid fall in tunnelling probability as $v$ increases. The latter has collected some of the energetic degrees of freedom but only a fraction of the number that are able to tunnel into the lower minimum. This behaviour provides further support for the presence of quantum tunnelling. There are other simple tests one could devise, and set-ups that probe different aspects of the physics, which will be the subject of future study. For example one could construct a potential with a small but thin extra barrier in front of the second well. Thermally excited transition would be greatly reduced  by such a barrier, while quantum transition would be virtually unaffected. A point we would like to emphasise however is the ease with which our framework allows one to formulate and address the question. 

\section{Conclusion}

Barrier penetration is a manifestly quantum mechanical property of a quantum field. While such tunnelling processes have been observed and studied in quantum mechanics and a selection of special quantum field theories realised in nature, for instance in some condensed matter systems, to our knowledge, such instanton processes have never been observed and experimentally studied in a freely chosen quantum field theory. 

For this purpose we outlined how to encode a quantum field theory as an Ising model and probe it experimentally. The quantum field is represented by a spin chain and each node corresponds to a qubit on a quantum annealer. After initialising the quantum field with a field value in the potential minimum, one can observe it settle into a quantum eigenstate characteristic of the potential profile imposed on the system.
In a second step we then modified the energy profile of the quantum annealer across its qubits, such that the quantum field was no longer in the global potential minimum, but in a false vacuum. We then measured the probability for the field to tunnel from the false to the true vacuum for various tunnelling times, vacuum displacements and potential profiles. It was then possible to compare the observed tunnelling probabilities with that predicted theoretically  by the WKB method. 

Thus a quantum annealer, as for example provided by D-Wave, is a genuine quantum system that, following our method, can be used as a quantum laboratory for general field theories. The complexity of the theory that can be studied in this laboratory is limited only by the number and connectivity of the qubits in the quantum annealer. This highly adaptive approach could therefore have far reaching implications for future studies of quantum field theories. As experimental measurements of the dynamical behaviour of field theories are entirely independent of theoretical calculations, they can be used to infer their properties without being limited by the availability of suitable perturbative or nonperturbative computational methods. Conversely, in the near future, measurements in such a quantum laboratory could be used to improve theoretical and computational methods conceptually. Furthermore it will enable the measurement and detailed study of previously unobserved quantum phenomena, involving solitons, instantons and so forth, that are relevant for field theories of interest in particle physics, condensed matter physics, quantum optics or cosmology.

\begin{acknowledgements} We are grateful to Nick Chancellor for discussions, and to D-Wave Systems and their community for technical support and suggestions during this work. 

\end{acknowledgements}

\bibliographystyle{mykp}
\bibliography{references}  

\end{document}